# Self-Diffusion of Iron in $L1_0$ FePd film - as revealed by reflectometric methods

D. G. Merkel, Sz. Sajti, F. Tanczikó, M. Major, Cs. Fetzer and L. Bottyán

KFKI Research Institute for Particle and Nuclear Physics,
P.O.B 49, H 1525, Budapest, Hungary

A. Kovács

MTA Research Institute for Technical Physics and Materials Science (MFA),
H 1525, P.O.B. 49 Budapest, Hungary

A. Rühm, J. Major

Max Planck Institute for Metals Research
Heisenbergstr. 3, 70569 Stuttgart, Germany

R. Rüffer

European Synchrotron Radiation Facility, BP 220, 38043 Grenoble CEDEX 9, France

## Abstract

$L1_0$ (CuAu(I)-type) ordered FePd as well as FePt and CoPt, due to their high magnetic anisotropy, are candidate materials for future ultra-high density magnetic recording. Atomic diffusion governs the structural relaxation and associated changes in the physical and magnetic properties in these alloys. Such processes involve diffusion paths of a few angstroms. In developing modern storage devices it is indispensable to understand these processes. To investigate self diffusion, an isotope sensitive technique is needed. Neutron reflectometry (NR) is a suitable non destructive method to study self-diffusion in a chemically homogenous isotope multilayer with diffusion lengths of the order of a few angstroms. Another recently emerged isotope sensitive method for studying self diffusion is Synchrotron Mössbauer Reflectometry (SMR)

Partially ordered and disordered isotope-periodic multilayer (ML) films of composition $[^{nat}Fe_{47}Pd_{53}(3\text{ nm})/\ ^{57}Fe_{47}Pd_{53}(2\text{ nm})]_{10}$ were epitaxially grown on MgO(001)by molecular beam epitaxy (MBE). Post grown annealing was performed in UHV chamber at varying temperature between 370 $^{o}$C and 570 $^{o}$C . The heat treatment time changed from 90 min to 1800 min.

According to our model, three different microstructural species were identified in the FePd layer and were related to the different hyperfine (hf) field distributions in the CEMS spectrum taken on the samples (top right figure). The ordered $L1_0$ structure attributed to the low hf fild with magnetisation pointing along the crystallographic *c* direction. The intermediate hf distribution with rando magnetic orientation represents the disordered fcc FePd component and the high hf field corresponds to an iron rich environment which is magnetically coupled to the matrix in what it is embedded.

By fitting the total diffusion parameters for all FePd samples obtained from NR and SMR evaluation we determined activation energies (Q) and preexponent factors ($D^0$) for the distinct environment in the FePd samples.

We obtained the highest activation energy (1.82±0.38 eV)and lowest preexponential factor((5.76±0.35)*$10^{-14}$ $m^2/s$) for the ordered $L1_0$ as oppose to the iron rich regions were the activation energy (1.48±0.26 eV) found to be lowest and the preexponential factor the highest ((1.01±0.61)*$10^{-13}$ $m^2/s$).

## Introduction

Due to the high perpendicular magnetic anisotropy, $L1_0$ FePd(Pt), CoPd(Pt) are in great interests for future high density magnetic recording media [1, 2, 3]. The structure of $L1_0$ corresponds to CuAu(I)-type structure which includes the alternation of Fe and Pd atomic planes. The formation and transformation of $L1_0$ phase in such films is controlled by short range (atomic scale) diffusion, therefore it is widely studied. In the case of chemical driving force the interlayer diffusion can be followed by X-Ray techniques but due to the concentration dependence of the diffusion coefficient, it cannot be described by only one Arrhenius equation, even if the annealing was performed on increasing temperatures. It is more advantageous if the intermixing is followed in only one crystallographic direction in an epitaxial, chemically homogenous alloy, where the diffusion doesn't have chemical driving force (besides the isotopic effect) and the isotopes of the same elements move according to the same diffusion coefficient. This method allows us to measure diffusion on a very short scale. It is very important in the study of metastable phases because through the extended diffusion, phase transition occurs hence the diffusion parameter changes irreversible. In previous measurements, however the authors don't mention, this effect can be observed. In the case of FeZrN the change of conversion electron Mössbauer (CEMS) spectrum after heat treatement indicates a phase transformation in the sample [4,]. In the case of FePt, according to the paper published by the same group, CEMS results show the rotation of the initially out of plane c-axis of FePt into the plane of the sample after increasing temperature annealing. According to Rennhofer et al. [5] the diffusion length differs in the crystallographic c- and a-b direction the diffusion constant changes not only due to the temperature. These systematic errors can only be eliminated (or take into account) if the phases or the orientation of the phases are followed by an independent measurement. CEMS is an ideal method where $^{57}$Fe containing samples are studied. For this reason in this study we have measured CEMS on all samples before and after annealing and the change of the structure was taken into account.

In former studies [4,5] the evaluation of synchrotron Mössbauer spectra (SMR) was performed by the normalization of the Bragg peak to the total reflexion peak but Andreeva showed [6] that ratio of the intensity of the total reflexion and Bragg peak is very sensitive to the hyperfine properties of the sample and also to the conditions of the sample during measurement (Θ and tilt angle setup respect to the incident beam). For this reason even if the sample is a periodic multilayer, it is necessary to know the properties of the sample (namely CEMS spectrum) for the determination of the diffusion profile from the SMR spectrum. The error of the peak ratio can be extremely large if a capping layer is used or Kiessig oscillation is present due to the buffer layer. For the proper evaluation the fitting of the whole curve is required and even better if the promt reflectivity spectrum and the delayed spectrum is fitted simultaneously.

**Experimental**

$^{nat}Fe_{47}Pd_{53}{}^{57}Fe_{47}Pd_{53}$ isotope periodic multilayers with different initial structure were grown on ×20×2 mm$^3$ MgO(001) substrates by molecular beam epitaxy (MBE) technique. Depending on the desired structure the substrate temperature was held at RT or 350$^o$C during growth process. For epitaxial deposition 3 nm Cr seed layer was applied on the substrate followed by 15 nm Pd buffer layer. Then the bi-layer sequence of 3 nm $^{nat}Fe_{47}Pd_{53}$ and 2 nm $^{57}Fe_{47}Pd_{53}$ was repeated ten times. All $^{nat}Fe_{47}Pd_{53}$ and $^{57}Fe_{47}Pd_{53}$ layers were prepared by co-evaporation of Fe and Pd at a rate of 0.0485 Å/s for both $^{57}$Fe and $^{nat}$Fe and 0.0682 Å/s for Pd. To avoid oxidation of Fe a 1 nm Pd layer was grown on top of the sample (capping layer). The base pressure of the evaporation chamber was 2×10$^{-10}$ mbar which raised 2.8×10$^{-9}$ mbar during the growth process. To achieve identical layer structure for the different samples the tooling factors were carefully calibrated by test films and the layer thicknesses and composition were determined by x-ray reflectometry and Rutherford Backscattering. During evaporation the growing rate was controlled by two independent quartz layer thickness monitors.

Depending on the experiment the samples were cut into eight (10×5×2 mm$^3$ for SMR) or four (10×10×2 mm$^3$ for NR) equal pieces. From each set one sample was left in the as deposited state and the rest were annealed between 297$^o$C and 530$^o$C for various times between 90 min and 1800 min. The heat treatment was carried out in UHV conditions. To get the real sample temperature, the sample surface was monitored with an infrared heat camera.

CEMS spectrum was recorded on all samples before and after heat treatment. The experiments were performed by using a $^{57}$Co(Rh) single line Mössbauer source with a homemade gas flow single-wire proportional counter operating with He gas with 4.7% CH4 extinction gas and at a bias voltage of 830±10 V.

SMR measurements were performed at the ID18 and ID22 beam line of the European Synchrotron Radiation Facility (ESRF, Grenoble, France) in 16 bunch mode. For sequential monochromatization of the beam to the 14.4 keV Mössbauer transition (($\lambda$=0.86025 Å) of $^{57}$Fe, a Si(111), then a Si(4.2.2)/Si(12.2.2) double-channel cut monochromator were used. Prompt (non-resonant) and $^{57}$Fe delayed (time integrated) nuclear resonant reflectograms were recorded on each sample.

NR curves were recorded at NREX$^{+}$ beamline in (FRM II), Garching. The instrument was used in simple reflectivity setup at a wavelength of 4.2Å using a supermirror filter for higher harmonics suppression. The reflected intensity was detected with a 200×200 mm$^2$ two-dimensional $^{3}$He wire chamber detector. At each reflection angle the counts were summed up in a carefully selected region (ROI) to determine the reflected intensity.

High angle X-ray diffraction was carried out using a D8 Discover type diffractometer (equipped with a Göbel-mirror on the primary side) in Bragg-Brentano geometry using Cu K$\alpha$ radiation ($\lambda$=1.5415 Å). In order to decrease beam divergence, 0.6 mm slits were used after the source and in front of the detector.

## Results and discussion

Fifty samples of layer composition isotope-periodic $^{57}$FePd/$^{nat}$FePd multilayer samples have been investigated after various retention time and annealing temperature. The interdiffusion of the adjacent isotope-periodic layers was determined by fitting the SMR and NR curves (examples of which are displayed as inset A, C, and B, D in Fig 3, repectively). In Figure 1 the fitted values of the diffusion coefficient are plotted as a function of the reciprocal temperature of annealing (Arrhenius plot).

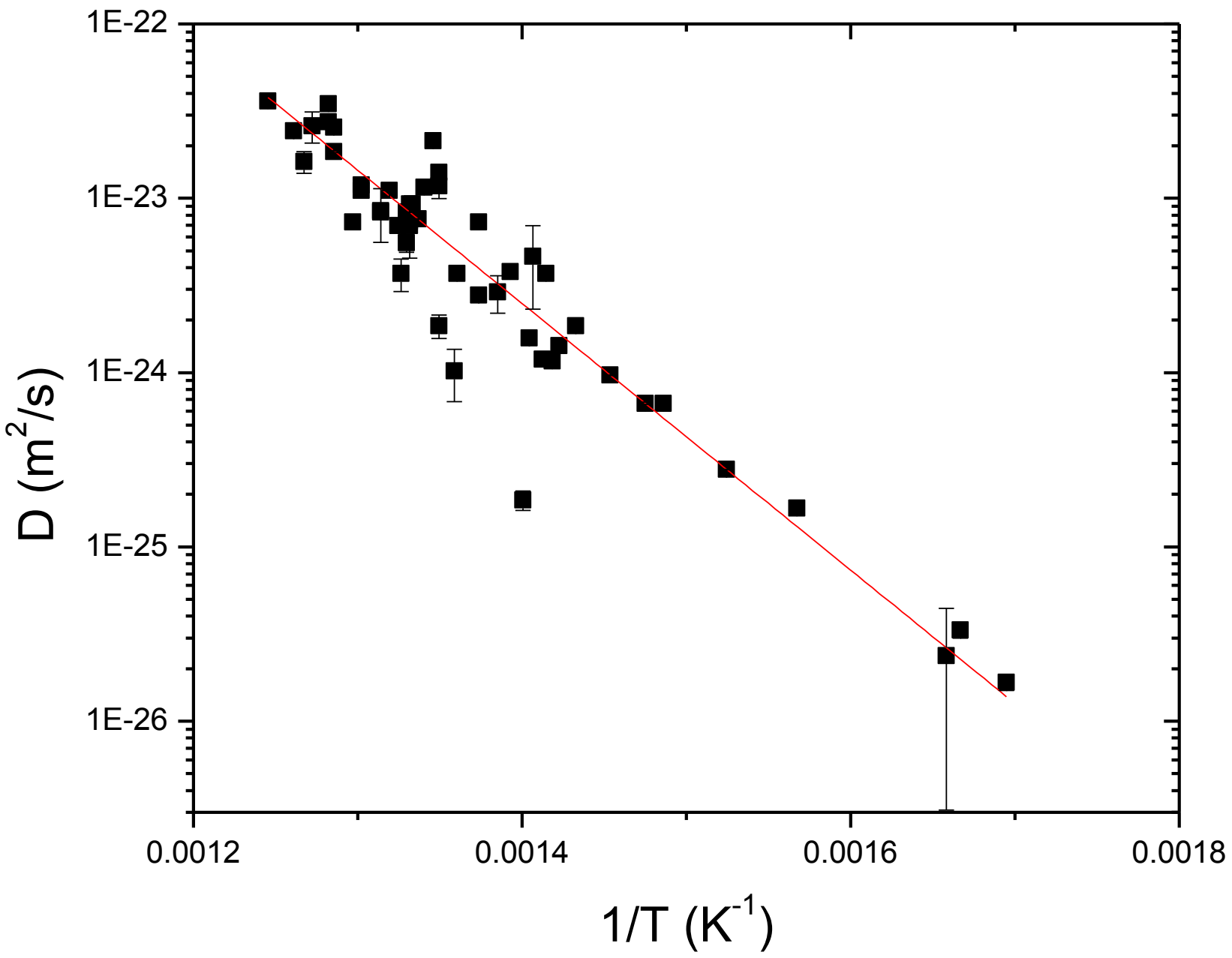

**Figure 1** Diffuzion coefficients for various annealing temperatures of FePd obtained from NR and SMR experiments

Obtaining activation energy and preexponal factor from fitting the points we get 1.51±0.31 eV for activation energy and and $(1.38\pm7.58)\times10^{-13}$ $m^2s^{-1}$) for pre-exponential factor. The points in Figure 1 show very diffuse feature. There are orders of magnitude (high above measurement error) scattering of the values at a given temperature. This behavior presumes that however the samples have the same nominal composition but there can be difference in the inner structure. To confirm this, CEMS measurements were taken on all samples. Three distinct microstructural species, the low hyperfine (hf) field, a large hf field, and an intermediate hf field species could be identified, which were attributed to the ordered $L1_0$, one iron rich nanoncluster phase and the disordered fcc structural units, respectively. On Figure 2 three selected Mössbauer spectra with their hyperfine field (hf) distributions are shown. On the top, the main component is $L1_0$ FePd (78%) with a 16% disordered fcc structure. The presence of iron rich clusters are almost neglectable (6%). In the middle the $L1_0$ and disordered environments gives almost identical contribution (~40%) to the spectrum and the ratio of high hf component is also significant (~20%). In the sample which is shown in bottom the high hf field enviroment is completely missing, only a little fraction of $L1_0$ (13%) FePd and mostly disordered fcc components are present.

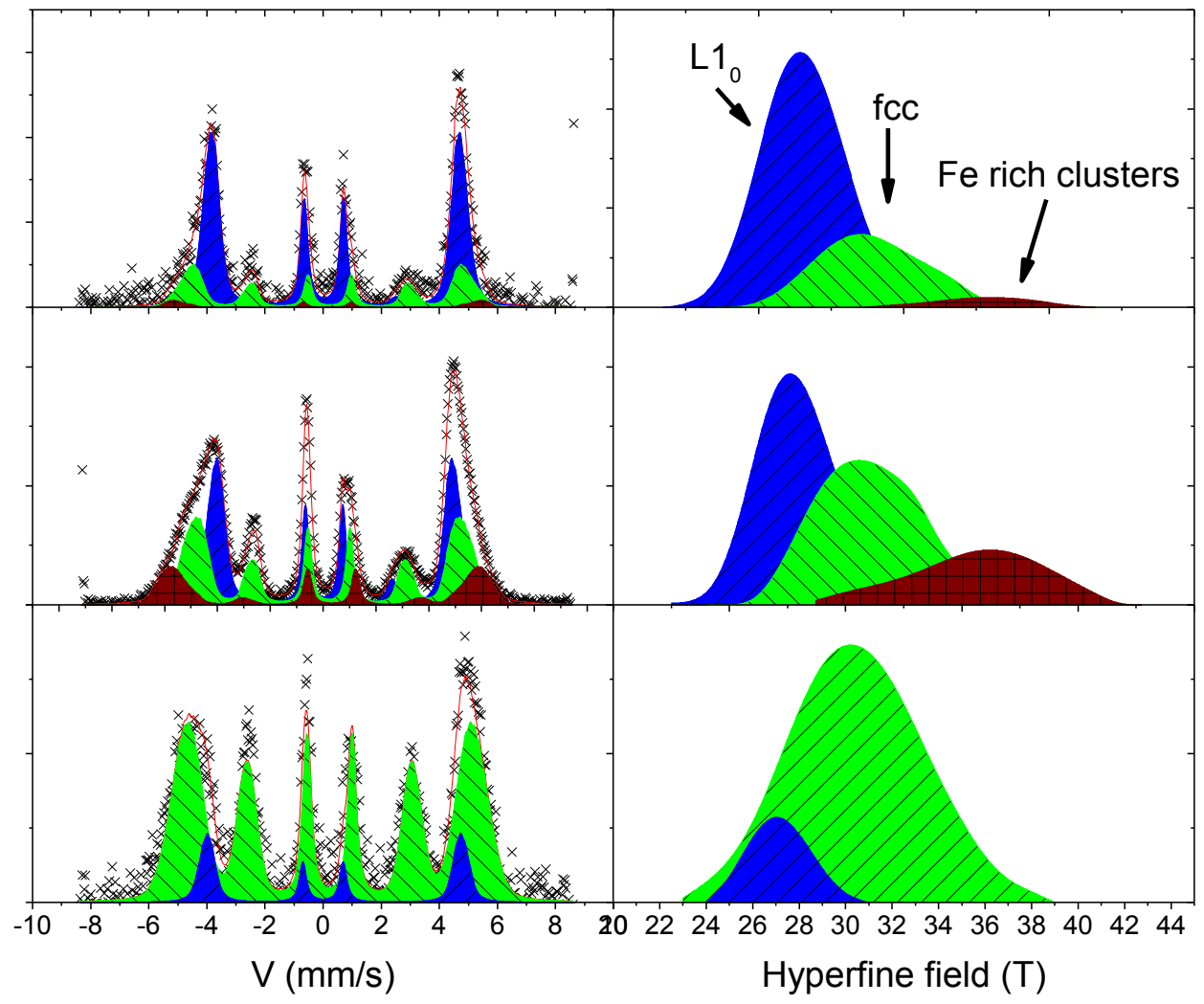

**Figure 2** Three selected Mössbauer spectra representing higly ordered, partially ordered and disordered FePd structures

These results indicate that the samples should be considered as inhomogeneous.

If we consider a 1D model where the diffusion coefficient (*D*) is piecewise constant, thus *D* takes an arbitrary positive value $D_i$ in volumes of arbitrary size, the diffusion of Fe in the multilayer, is, like in a random alloy, a weighed sum of the individual diffusion coefficients of iron in the different Fe-environments, , i.e. the total diffusion can be written as follows.

$$D = x_{L1_0} D_{L1_0} + x_{fcc} D_{fcc} + x_{Fe} D_{Fe}$$

Where $x_{\mathrm{L10}}$, $x_{\mathrm{fcc}}$,$x_{\mathrm{Fe}}$ and $D_{\mathrm{L10}}$, $D_{\mathrm{fcc}}$, $D_{\mathrm{Fe}}$ are the ratios and the effective diffusion parameters of the ordered $L1_0$, disordered fcc and iron-rich phases, respectively. Expressing the above equation we get the following form:

$$D(T) = x_{L1_0} D^o_{L1_0} e^{-\frac{Q_{L10}}{kT}} + x_{fcc} D^o_{fcc} e^{-\frac{Q_{fcc}}{kT}} + x_{Fe} D^o_{Fe} e^{-\frac{Q_{Fe}}{kT}}$$

By fitting the diffusion values for all 49 samples obtained from NR and SMR evaluation we determined activation energies ($Q$) and preexponent factors ($D^0$) for the distinct environments in the FePd samples. On Figure 3 the measured diffusion and the result of fitting is shown. For selected samples the corresponding NR or SMR spectrum is plotted. On the top of the figure the composition of the samples are presented.

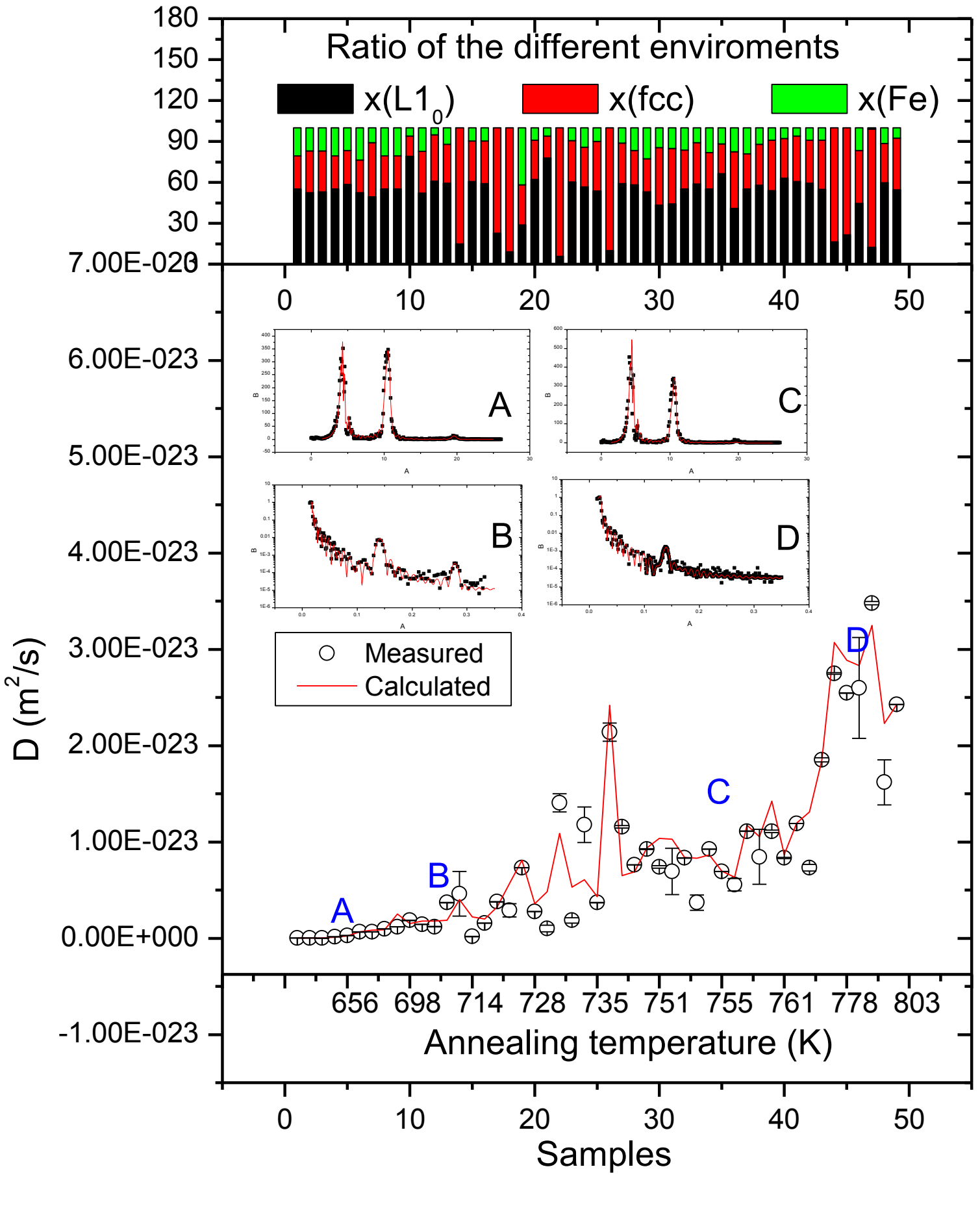

**Figure 3** The measured and fitted diffusion coefficients for all FePd samples (bottom). On the top of the figure, the relative fraction of the different environments in a given sample is shown. Selected NR and SMR spectra after different annealing temperature are shown in the insets

From fitting the pre-exponential factors and the activation energies founded to be $D^0_{L10}$=(5.76±0.35)*$10^{-14}$ $m^2s^{-1}$, $Q_{L10}$=1.82±0.38 eV, $D^0_{fcc}$=(1.32±0.45)*$10^{-13}$ $m^2s^{-1}$ $Q_{fcc}$=1.48±0.26 eV, $D^0_{Fe}$=(1.01±0.61)*$10^{-13}$ $m^2s^{-1}$, $Q_{Fe}$=1.39±0.23 eV for $L1_0$ ordered, fcc disordered FePd and for iron rich regions, respectively. Note, that these values are in the crystallographic *c* direction which is perpendicular to the sample surface. According to these results the diffusion in the $^{57}$FePd/$^{nat}$FePd is controlled by the disordered fcc and the iron rich regions. In the ordered $L1_0$ regime the atomic movement is almost blocked compared to the other structures.

**Summary**

Partially ordered and disordered isotope-periodic multilayer (ML) films of composition [$^{nat}Fe_{47}Pd_{53}$(3 nm)/ $^{57}Fe_{47}Pd_{53}$(2 nm)]$_{10}$ were epitaxially grown by molecular beam epitaxy (MBE) on Pd(001)(15 nm) buffer and Cr(3nm) seed layer on MgO(001). A Pd(1nm) capping layer was used to prevent oxidation. Post grown annealing was performed in UHV chamber (˜$10^{-10}$mbar) at varying temperature between 370 $^o$C and 570 $^o$C . The annealing time varied from 90 min to 1800 min. During annealing the surface temperature of the sample was monitored by an infrated thermometer.

According to our model, three different microstructural species were identified in the FePd layer and were related to the different hyperfine (hf) field distributions in the CEMS spectrum taken on the samples. The ordered $L1_0$ structure attributed to the low hf fild with magnetisation pointing along the crystallographic *c* direction. The intermediate hf distribution with rando magnetic orientation represents the disordered fcc FePd component and the high hf field corresponds to an iron rich environment which is magnetically coupled to the matrix in what it is embedded.

In our model the diffusion (*D*) piecewise constant, thus *D* takes an arbitrary positive value $D_i$ in volumes of arbitrary size, so the diffusion of Fe in the multilayer, is, like in a random alloy, a weighed sums of the individual diffusion of iron of the different Fe-environments.

By fitting the total diffusion parameters for each FePd samples obtained from NR and SMR evaluation we determined activation energies (Q) and preexponent factors ($D^0$) for the distinct environment in the FePd samples.

We obtained the highest activation energy (1.82±0.38 eV)and lowest preexponential factor(($(5.76\pm0.35)*10^{-14}$ $m^2/s$) for the ordered $L1_0$ as oppose to the iron rich regions were the activation energy (1.48±0.26 eV) found to be lowest and the preexponential factor the highest (($(1.01\pm0.61)*10^{-13}$ $m^2/s$).

**Acknowledgement**
This work was supported by the Hungarian National Science Fund (OTKA) and by the National Office for Research and Technology of Hungary under contract K 62272 and NAP-VENEUS'05.